\documentclass[amsmath,amssymb,aps,prl,twocolumn,superscriptaddress,showpacs,10pt]{revtex4-1}
\usepackage{mathbbol,amsmath,amsfonts,amssymb,bm,color,ulem}
\usepackage{graphicx,psfrag,array,braket,mathtools}
\usepackage[resetlabels,labeled]{multibib}
\newcites{S}{Supp}

\newcommand{\comment}[1]{}
\newcommand{\Eq}[1]{Eq.~(\ref{#1})}

\newcommand{\fr}{\mathfrak{r}}
\newcommand{\sfr}{{\sf root}}

\def\be{\begin{equation}}
\def\ee{\end{equation}}
\begin{document}
\title{Local Two-Body Parent Hamiltonians for the Entire Jain Sequence}
\author{Sumanta Bandyopadhyay}
\affiliation{Department of Physics, Washington University, St. Louis, MO 63130, USA}
\affiliation{Nordita, KTH Royal Institute of Technology and Stockholm University, Roslagstullsbacken 23, SE-106 91 Stockholm, Sweden}
\author{Gerardo Ortiz}
\affiliation{Department of Physics, Indiana University, Bloomington, IN 47405-7105,USA}
\affiliation{Department of Physics, Washington University, St. Louis, MO 63130, USA}
\author{Zohar Nussinov}
\author{Alexander Seidel}
\affiliation{Department of Physics, Washington University, St. Louis, MO 63130, USA}
\date{\today}
\begin{abstract}
Using an algebra of second-quantized operators, we develop local two-body parent Hamiltonians for all unprojected Jain states at filling factor $n/(2n\sf p+1)$, with integer $n$ and (half-)integer {\sf p}. We rigorously establish that these states are uniquely stabilized and that zero mode counting reproduces mode counting in the associated edge conformal field theory. We further establish the organizing ``entangled Pauli principle'' behind the resulting zero mode paradigm, and unveil an emergent SU($n$) symmetry characteristic of the fixed point physics of the Jain quantum Hall fluid.
\end{abstract}
\maketitle
\textit{Introduction.$-$}
The fractional quantum Hall (FQH) effect enjoys a unique position in strongly correlated electron physics both as a fascinating physical effect \cite{STGexperiment} as well as a central juncture for the percolation of ideas between correlated electron physics and other areas of theoretical and mathematical physics. Originally, the success of the field owes much to construction principles for variational wave functions \cite{Laughlin, Jain, Halperin, MR,Milovanovic} and associated ideas to connect the latter to effective field { theories} \cite{MR, Wen90,fradkinLopez,murthy1998field}. In our opinion, the intimacy of the connection between microscopics and effective quantum field theory that is achievable in this field is, in some cases, essentially unparalleled. This is the case when the construction of a parent Hamiltonian \cite{haldane_hierarchy, TK, MR, Halperin} is possible that falls into what we term the ``zero mode (ZM) paradigm'': The zero (energy) mode space of a positive { semidefinite} Hamiltonian is composed of an incompressible state as well as edge or quasihole excitations, where the counting of ZMs in each angular momentum sector (relative to the incompressible state) precisely matches \cite{Milovanovic,read09conformal,Readpaper} the mode counting in the conformal edge theory. This then unambiguously points to the edge conformal field theory associated to the state, and, thanks to bulk-edge correspondence, all universal physics are then essentially fixed through exact properties of the microscopic Hamiltonian.

While a considerable number of these very special Hamiltonians exist, they are absent for many phases that are of central importance to the theory of the Hall effect. The latter include almost all phases described by Jain composite fermion (CF) states \cite{JainCF, Jain1, Jain2, Jain3}, which are key to the understanding of the physics at Landau level (LL) filling  factor $\nu<1$. While some Hamiltonians have been proposed for (non-Laughlin) Jain-type states \cite{MacRe, Kivelson, Sreejith_Jain}, a ZM paradigm has only been established at 2/5 in Ref. \cite{Chen17} ({ for other more exotic parton states in Ref. \cite{Sumanta18}}). There, some of us have argued that such a paradigm is possible in principle only for {\it unprojected} Jain states, which are well known to be in the same phase as their projected counterparts \cite{MacRe,JainAdvPhys}. In this case, traditional first-quantized construction principles for parent Hamiltonians face unusual challenges. The latter seek to enforce ``analytic clustering properties''\cite{analclus00,analclus01,analclus02} in the few-body density matrices of ZMs \cite{haldane_hierarchy, TK, MR, Halperin,Wenpatternsofzeros1, Wenpatternsofzeros2}. Indeed, unprojected Jain states generally have a zero of order 2{\sf p}+1, { with (half-) integer {\sf p} $\ge 0$}, when two particles meet at the same point. However, enforcing just this (2{\sf p}+1)-clustering property will generally lead to more exotic ``parton'' states \cite{Jain, Wen90, Wen90_1} as the incompressible ground states when more than $n=2$ LLs are present \cite{Sumanta18, Mostafatobepublished}. Actually, the (2{\sf p}+1)-clustering property comes from a purely holomorphic factor of the wave function while an antiholomorphic dependence is also present. This additional information is not straightforwardly enforced  through a local Hermitian few-body interaction.

In this work, we solve this problem for all Jain CF states at filling factors $n/(2n\sf p+1)$, with integer { $n\ge1$}. We utilize a recently developed operator formalism \cite{Chen18} that describes CFs as second-quantized objects in  Fock space. This leads to an algebraic construction of the parent Hamiltonian that represents a radical departure from the traditional constructions principles described above, and fully embraces the ``guiding-center-only'' approach to FQH physics that has recently been influential  \cite{Haldane11,ortiz}. Our results have further  important ramifications for the theory of frustration free lattice Hamiltonians, in that we establish a framework where these become tractable even if interactions are {\it not} strictly short-ranged in generalized lattice coordinates. A close connection with the recently celebrated matrix-product structure of many FQH states  \cite{dubail, zaletel, estienne} is anticipated, though we leave details for future work \cite{MAtheustobepublished}.

\textit{Composite fermions and Zero modes.$-$} The unprojected (mixed-LL) Jain state at filling factor $\nu=n/(2 n \sf p+1)$ \cite{note11} can be defined in disk geometry as 
\be\label{JS}
\Psi_{n,\sf p}(N)=\prod_{1\leq i<j\leq N} (z_i-z_j)^{2\sf p} \Phi_n(N),
\ee 
where $\Phi_n(N)$ denotes an integer quantum Hall (IQH) state of $N$ particles in $n$ LLs, and the $z_i=x_i+{\rm i}y_i$, $\bar z_i=x_i-{\rm i}y_i$ are the particles' complex coordinates.  $\Phi_n(N)$ is by definition a state of ``densest'' possible electron configuration for given $n$ and $N$, where ambiguities at the edge may arise for certain $N$ that we will  resolve below.

 Equation \eqref{JS} clearly has a ``clustering property,'' where the wave function has a (2{\sf p}+1)th order zero when two particles converge to the same point. However, only for $n=2$ \cite{Chen17, Sumanta18} does \Eq{JS} represent the densest ({ largest} filling factor) wave function(s) having this property. Related to that, for $n=2$, ${\sf p}=1$  there is the aforementioned,  well-documented parent Hamiltonian satisfying the zero  mode paradigm. To solve the general problem, we turn to an alternative characterization given by some of us \cite{Chen18} in terms of an algebra of second-quantized operators, which can be understood as ``ZM generators''. We begin by summarizing the nuts and bolts of this formalism. 

In first-quantization, an orbital $\phi_{m,\ell}$ in the { $(m+1)$th} LL, $m=0,1\dotsc$, with angular momentum $\ell$ is a superposition of monomials of the
form $\mu_{a,\ell}=\bar z^{a} z^{\ell+a}$ with $0 \leq a\leq m$. (We omit obligatory Gaussian factors.) Higher LL many-body wave functions such as \Eq{JS} may be expanded in $\mu_{a,\ell}$, adorned with additional particle indices. A significant advantage of the first-quantized presentation is the fact that this expansion is essentially {\it geometry independent}, assuming that we limit ourselves to zero genus geometries (disk, cylinder, sphere) \cite{ortiz}. This is so since there is a one-to-one correspondence between the wave functions in these geometries, once $\bar z$, $z$ (for the disk) are replaced with suitable functions of coordinates respecting the boundary conditions of the respective geometries. In other words, variational wave functions such as \Eq{JS} are described by the same polynomials in the genus 0 geometries. To obtain a manifestly geometry independent language, and to the extent that the successful construction of a parent Hamiltonian is a direct consequence of the underlying polynomial structure, however complicated, it proves advantageous to make the monomials $\mu_{a,\ell}$ the essential degrees of freedom of the second-quantized formalism also. For fixed $a$, we think of these orbitals as constituting a ``$\Lambda$ level'' { ($\Lambda$L)}. We thus introduce pseudofermion \cite{pseudofermion} operators $\tilde c_{a,\ell}^{\;}$, $\tilde c_{a,\ell}^\ast$ satisfying canonical anticommutation relations
\be\label{cac}
\{ \tilde c_{a,\ell}^{\;},  \tilde c_{a',\ell'}^\ast\}=\delta_{a,a'}\delta_{\ell,\ell'},
\ee
where $\tilde c_{a,\ell}^\ast$ { creates} an electron in the orbital $\mu_{a,\ell}$. These orbitals are {\it not} normalized or orthogonal (for fixed $\ell$), and hence $\tilde  c_{a,\ell}^\ast$ and $\tilde  c_{a,\ell}$ are not Hermitian conjugates, but this will present no obstacle in the following. If desired, at the end we may always return to the canonical creation and annihilation operators $c_{m,\ell}^{\;}$, $c_{m,\ell}^\dagger$ of the orbitals $\phi_{m,\ell}$ via
\be\label{Amatrix}
    c_{m,\ell}^\dagger =\sum_{a=0}^{m}  A^{-1}(\ell)_{m,a} \tilde c_{a,\ell}^\ast \,,\;\; c_{m,\ell}= \sum_{a=m}^{n-1} \tilde c_{a,\ell} A(\ell)_{a,m}\,.
\ee
The (real, lower-triangular) matrix $A(\ell)$ is the only geometry-dependent aspect of this formalism. It is given in \cite{supplement} 
for the disk/cylinder geometries.

The considerable advantage of the second-quantized formalism \cite{ortiz}, especially for multiple LLs,
lies in the fact that it gives us control over an algebra of ``ZM generators'' we arguably do not have in first-quantization. It is also much more conducive to recursive schemes in particle number which we will now heavily pursue. To this end we introduce the following operators, which we will think of as ZM generators in a sense to be made precise:
\be\label{pkab}
   \hat p_k^{a,b} =\sum_{\ell}\tilde{c}^{*}_{a,\ell+k}\tilde{c}_{b,\ell}^{\;}\;   \mbox{($a\geq b\!-\! k$ for disk geometry),}
\ee
and which is a generalization of the operator ${\cal O}_d$ introduced in Ref. [\onlinecite{ortiz}] for multiple LLs.

The operators in \Eq{pkab} generate an algebra (via taking sums and/or products) that we denote by $\mathcal{Z}$. The significance of this algebra is manifold \cite{Chen18}. It allows for a definition of CF states recursive in particle number, quite distinct from the recently fashionable matrix-product presentation of FQH states \cite{zaletel, estienne}, but it is in essence a generalization of Read's expression of the Laughlin state through a nonlocal order parameter \cite{ReadOP,PhysRevB.91.085115}. Indeed, the algebra allows for a microscopic definition \cite{Chen18} of a complete set of order parameters for CF states. In the present context, it will turn out that the algebra $\mathcal{Z}$ generates all possible ZMs when acting on the incompressible ground state. In that sense they are related to the first-quantized formalism discussed by Stone \cite{stonepaper} for the Laughlin state, possible there because $\sum_{a=0}^{n-1} \hat p_k^{a,a}$ [which, for $n=1$ LL, is really all \Eq{pkab} boils down to] has a simple first-quantized interpretation: It multiplies many-body wave functions with power-sum symmetric polynomials $p_z=\sum z_i^k$ \cite{ortiz,PhysRevB.91.085115}. For multiple LLs, however, we need the full set $\hat p_k^{a,b}$, which does {\it not} 
have a straightforward first-quantized interpretation \cite{Chen18, Mostafatobepublished}.

Consider now \Eq{JS}.  To resolve the ``edge ambiguity'' mentioned above, we define the Slater determinant
by successively filling the state $\mu_{a,\ell}$ with lowest available $\ell+a$ that has lowest not-yet-occupied $a$.
We seek to establish a parent Hamiltonian such that \Eq{JS}, which we now also suitably write $\ket{\Psi_{n,{\sf p},N}}$,
is a ZM of this Hamiltonian. Since general ZMs will describe edge excitations and, deeper in the bulk,
quasihole excitations \cite{justify}, one has the intuition \cite{ReadOP} that $\tilde c_{a,\ell}\ket{\Psi_{n,{\sf p},N}}$, is also a ZM 
of the Hamiltonian, namely, one describing a cluster of quasiholes of total charge 1 inserted into $\ket{\Psi_{n,{\sf p},N}}$. Anticipating that this is so, then, with the properties of the
$\hat p_k^{a,b}$ as advertised,
we must be able to interpret this
as a ZM generated by some combination of $\hat p_k^{a,b}$ on top of the reference state $\ket{\Psi_{n,{\sf p},N-1}}$, or
\be\label{1hole}
\tilde c_{a,\ell}\ket{\Psi_{n,{\sf p},N}} = \hat Z_{n,{\sf p},N,a,\ell} \ket{\Psi_{n,{\sf p},N-1}}\,,
\ee
where $\hat Z_{n,{\sf p},N,a,\ell}$ is a suitable element of the algebra $\mathcal{Z}$. Indeed, the relation between $\hat Z_{n,{\sf p},N,a,\ell}$ and the generators \eqref{pkab} was made explicit in Ref. \cite{Chen18}, but will not be needed in the following. 

\textit{Parent Hamiltonian for composite fermions.$-$} 
We are now ready to present the following Hamiltonian,
\begin{eqnarray}\label{HMn}
  H_{n,\sf p}&=& \sum\limits_{ J, r, a, b}
  E^{r}_{a,b,J} T^{r\dagger }_{a,b,J}T^r_{a,b,J},\\    
 T^r_{a,b,J}&=&\sum_{x} x^r \tilde{c}_{a,J+x}\tilde{c}_{b,J-x} \, ,
\end{eqnarray}
where $J$ runs over half-integer values with $J\geq -n$, $0\leq r<2\sf p$, $0\leq a\leq b<n$.  The ${T^{r\dagger}_{a,b,J}} T^r_{a,b,J}$ \cite{comm2} are suitable generalizations of pseudopotentials, whose relation to Haldane pseudopotentials for $n=1$ was discussed in Ref. \cite{ortiz}. The $E^r_{a,b,J}$ are positive constants and may be used to enforce desirable spatial symmetries. We show in Ref. \cite{supplement} that positive $E^r_{a,b,J}$ can always be chosen so as to render the resulting Hamiltonian local. The $T^r_{a,b,J}$ may also be replaced with new linearly independent combinations without affecting the ZM space. It is worth noting that the absence of a kinetic energy splitting between the first $n$ LLs is a feature that is realized in certain stackings of multilayer graphene \cite{MLGraphene1,MLGraphene2,MLGraphene3}.

For fermions, $ T^r_{a,b,J}$ vanishes for even $r$ and $a=b$, giving ${\sf p} n^2$ different pseudopotentials at each pair-angular-momentum $2J$. Assuming disk geometry, we use the convention $\tilde c_{a,\ell}\equiv$ 0 for $a+\ell<$ 0. A key observation is that the operators $ T^r_{a,b,J}$ and $\hat p^{a,b}_k$ satisfy the following commutation relation:
\begin{eqnarray}
\label{comm}\nonumber
&& \hspace*{-0.8pt}[ T^r_{a,b,J}, \hat p^{a',b'}_k ]=\sum_{{\tilde r}=0}^r\binom{r}{{\tilde r}}\left(\frac{k}{2}\right)^{r-{\tilde r}} \times \\&&\left((-1)^{r-{\tilde r}}T^{{\tilde r}}_{a,b',J- k/2}\delta_{b,a'}+T^{{\tilde r}}_{b',b,J-k/2}\delta_{a,a'}\right).
\end{eqnarray}
This justifies the notion that the $\hat p^{a,b}_k$ are ZM generators: 
The condition for $\ket{\psi}$ to be a ZM of the { positive semidefinite} Hamiltonian \eqref{HMn} reads ${T}^r_{a,b,J}\ket{\psi}=0$ for all $r$, $J$, $a$, $b$. The commutator \eqref{comm} thus clearly vanishes within the ZM subspace. Therefore, any $\hat p^{a,b}_k$ acting on $\ket{\psi}$ immediately generates another ZM, with angular momentum increased by $k$. In the following,
we first wish to (i) establish that the Jain state $\ket{\Psi_{n,{\sf p},N}}$ is a ZM of \Eq{HMn}, and (ii) find all ZMs of \Eq{HMn}.

We achieve these goals via a radical departure from established paradigms, i.e., not paying attention whatsoever to analytic clustering properties. We will do so by utilizing the properties of the second-quantized operator algebras given above and in the following. For part (i), we give a simple induction proof in  $N$ which extends that of \cite{Chen14}. We give the induction step first, assuming that $\ket{\Psi_{n,{\sf p},N-1}}$ is known to be a ZM. One easily verifies $ T^r_{a,b,J}=\frac 12 \sum_{\tilde{a},\ell} [T^r_{a,b,J},\tilde c_{{\tilde a},\ell}^*]\tilde c_{{\tilde a},\ell}^{\;}\,.$ We apply this to $\ket{\Psi_{n,{\sf p},N}}$. Using \Eq{1hole} together with the fact that $\hat Z_{n,{\sf p},N,a,\ell}$ is a ZM generator, i.e., $T^r_{a,b,J}$ annihilates \Eq{1hole},
and that $\sum_{a,\ell}\tilde c_{a,\ell}^*\tilde c_{a,\ell}^{\;}$ gives the total particle number $N$, yields $T^r_{a,b,J}\ket{\Psi_{n,{\sf p},N}}= \frac{N}{2}T^r_{a,b,J}\ket{\Psi_{n,{\sf p},N}} $, or $T^r_{a,b,J}\ket{\Psi_{n,{\sf p},N}} =0$ for $N>2$. So far, the only special property of the { $T^r$ operators} ($0\leq r<2 \sf p$) that we have used is that $\hat Z_{n,{\sf p},N,a,\ell}$ is a ZM generator as defined above.
All that is left to do is to establish an induction beginning for $N=2$.
Indeed, the $N=2$ state in the class of states $\ket{\Psi_{n,{\sf p},N}}$  has the wave function $(z_1-z_2)^{2\sf p}(\bar{z}_1-\bar{z}_2)$, or, in second-quantization, 
\be
\begin{split}
|\Psi_{n,{\sf p},2}\rangle=\sum_{j}(-1)^j \binom{2\sf p}{j}\tilde{c}^*_{1,j-1}\tilde{c}^{*}_{0,2{\sf p}-j}\,|0\rangle.
\end{split}
\ee
This has angular momentum $2{\cal J}=2{\sf p}-1$, and the only { $T^r$ operators} that could possibly not annihilate the state are of the form $T^r_{0,1,{\cal J}}$. Acting with these operators produces
\be\label{induct}
\sum_{j} \left(j-{\sf p}-\frac 12\right)^r (-1)^{j}\binom{2{\sf p}}{j}|0\rangle=0,
\ee
for $r<2\sf p$, since indeed \cite{Ruiz} $\sum_{j=0}^{2\sf p} (-1)^j \binom{2\sf p}{j}(x-j)^{2\sf p}=(2\sf p)!$ independent of $x$, such that taking $x$ derivatives implies \Eq{induct}. 
 
\textit{Entangled Pauli principle (EPP).$-$} Having now established that the Jain state $\ket{\Psi_{n,{\sf p},N}}$ is a ground state of the Hamiltonian $H_{n,\sf p}$, \Eq{HMn}, we seek to understand the full ZM space of these Hamiltonians. This will, in particular establish the {\it densest} ZM(s) of this Hamiltonian, whose existence is generally taken as the hallmark of incompressibility. The key to obtaining such results for Hamiltonians of the form \eqref{HMn} lies in the fact that there is a now well-established \cite{ortiz, Chen14, Chen17, Sumanta18} general method to derive necessary conditions, in the form of EPPs \cite{Sumanta18}, on the ``root states'' for ZMs of such Hamiltonians. These root states encode the DNA of the incompressible fluids. Using these techniques we now establish that a complete set of ZMs for $H_{n,\sf p}$  is of the form \eqref{JS}, with the IQH state $\Phi_n$ replaced by ${\cal S}_n$, a generic Slater determinant with definite occupancies in $n$ Landau-/$\Lambda$Ls. That indeed such states are ZMs follows easily from the fact that the $\hat p_k^{a,b}$ are ZM generators, together with the convenient property that they commute \cite{Chen18} with the Laughlin-Jastrow flux-attachment operator. Acting on \Eq{JS}, the $\hat p_k^{a,b}$ may thus be thought of as acting {\it directly} on the IQH factor  $\Phi_n$, thus, on $\Lambda$L degrees of freedom. It is easy to see that any ${\cal S}_n$ can be generated out of $\Phi_n$ by acting with appropriate products of $\hat p_k^{a,b}$'s.
 
\begin{figure}
    \centering
    \includegraphics[scale=.36]{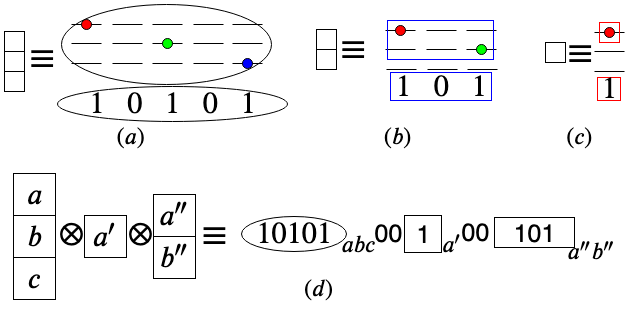}
    \caption{Graphic representation of the clusters  emerging at root level  $\ket{\Psi}_{\sf root}$.  ($a$)-($c$)
    show the individual building blocks for root states
    assuming $n=3, {\sf p}=1$, such that the underlying group structure is SU(3) (see text). The oval ($a$) denotes a singlet.
    ($d$) represents a sample root structure. Clusters generally consist of up to $n$ particles at distance $2{\sf p}$, totally anti-symmetric in $\Lambda$L-indices. Clusters are further characterized by the 
    $\Lambda$L-indices that are occupied, and must be mutually separated by at least $2{\sf p}+1$ orbitals.
    \label{fig1}}
\end{figure}

 Consider now the expansion of any ZM $\ket{\psi}$ into $\Lambda$L Slater determinants:
 \be\label{Slater}
 \ket{\psi}=\sum C_{(a_1,\fr_1),\dotsc, (a_N,\fr_N)} \tilde d^\ast_{a_1,\fr_1}\dotsc
 \tilde d^\ast_{a_N,\fr_N}\ket{0}\,,
 \ee
 %Here and in the following, we assume $r_i\leq r_{i+1}$ unless otherwise stated. 
 where we introduce 
 $\tilde d_{a,\fr} =\tilde c_{a,\fr-a}$, $\tilde d_{a,\fr}^\ast =\tilde c_{a,\fr-a}^\ast$, with labels that refer to a ``pseudo-guiding-center''  $\mathfrak{R}=\sum_{a,\fr} \fr \, \tilde d_{a,\fr}^\ast \tilde d_{a,\fr}$ \cite{specialize}.
 { This renders $\fr$ to be non-negative, just as $a$}.
 We define terms in the expansion \eqref{Slater}
 as ``nonexpandable'' \cite{ortiz} if 
 the action with every possible ``expansion'' operator of the form
 $\tilde d_{a_1',\fr_1-x}^\ast \tilde d_{a_2',\fr_2+x}^\ast \tilde
 d_{a_1,\fr_1}^{\;}\tilde d_{a_2,\fr_2}^{\;}$, $\fr_1\leq \fr_2$, $x>0$, leads to a term with zero coefficient.
The root state of $\ket{\psi}$, $\ket{\psi}_{\sf root}$, is now defined as that part of the expansion \eqref{Slater} consisting only of nonexpandable terms. $\ket{\psi}_{\sf root}$ 
so defined is necessarily nonvanishing due to the finite dimensionality of the subspace of given  $\mathfrak{R}$ \cite{Sumanta18,footnote1}. { As shown} in Ref. \cite{supplement}, $\ket{\psi}_{\sf root}$ is subject to the following EPP. i) The $\fr$ values of any two occupied single-particle states differ at least by $2\sf p$. ii) If they differ precisely by $2\sf p$, the root-level coefficients have the following antisymmetry property in $\Lambda$L indices:
\be
C_{\dotsc(a_i,\fr_i),(a_{i+1},\fr_i+2\sf p)\dotsc} = -C_{\dotsc(a_{i+1},\fr_i),(a_{i},\fr_i+2\sf p)\dotsc}\,.
\ee
%As a consequence, iii) for disk geometry, we have the following additional boundary condition: If a cluster of
%particles separated by distance 2$\sf p$ involves any orbital with negative angular momentum index $m<0$, the cluster's $\Lambda L$ indices $a_i$ are limited to the range $-m\leq a_i <n$. 
As in many known examples, the EPP immediately reveals the densest possible filling factor at which ZMs of the model \eqref{HMn} may exist. To this end, it is useful to translate the EPP into a language of SU($n$)-spins, where each spin carries the fundamental representation. We may think of the $\Lambda$L-index of a particle as an SU($n$)-index, and of its $\fr$-index as the position in a one-dimensional lattice. Then, permissible root states must be (linear combinations of) product states associated with certain clusters, each cluster containing up to $n$ particles. Within each cluster, particles are 2{\sf p} sites apart, and the ``spin'' wave function of each cluster is totally anti-symmetric. This renders the largest possible cluster an ``SU($n$)-singlet'' of $n$ spins [Fig. \ref{fig1}a], and clusters must be separated by at least 2{\sf p}+1 sites. It is easy to see that the densest possible root state is just a product of such clusters at a filling factor of $n$/($2n\sf p$+1).
 There are thus no ZMs whose filling factor can exceed this value in the thermodynamic limit, and the corresponding Jain state just satisfies this bound. One can, more generally, show \cite{supplement} that the number of possible root states sets an upper bound for the number of ZMs present in each angular-momentum-/$\mathfrak{R}$ sector. A  state counting argument shows, in turn, that the number of CF states of the form (Jastrow factor)$\times {\cal S}_n$ precisely { saturates} this bound \cite{supplement}. Therefore, such CF states form a complete set of ZMs of Eq. \eqref{HMn}. It is further easy to see that the counting of such CF states in a given angular momentum sector (relative to a minimum angular momentum CF state) coincides with the number of modes in the expected edge theory of $n$ branches of chiral fermions or bosons. This is pleasingly consistent  with the fact that these ZMs are all generated by the application of the bosonic ``density modes'' \eqref{pkab} on the reference state \eqref{JS}, and that these modes have the simple action on the Slater parts of CF states stated above. The Hamiltonians constructed here are thus  true representatives of the ZM paradigm discussed initially.
 Detailed numerical verification of this result is reported in Ref. \cite{supplement}.

\textit{Emergent SU($n$) symmetry.$-$}
In essence, the above establishes that root states, $\ket{\Psi}_{\sf root}$,  come as products of representations of SU($n$). Indeed, an underlying SU($n$)-symmetry is present not only at root level, but is an emergent property of the full ZM space. To make this symmetry readily visible, we write the commutation relations of the zero-mode generators \cite{Chen18}:
\be\label{Lie}[\hat p^{a,b}_k,\hat p^{b',a'}_{k'}]=\delta_{b,b'}\hat p^{a,a'}_{k+k'}-\delta_{a,a'}\hat p^{b',b}_{k+k'}. \ee
In a cylindrical  geometry, where there is no constraint on the subscript $k$, the above commutator is just the loop-algebra of SU($n$). In particular, for $k=k'=0$, we recover the algebra of SU($n$) itself \cite{point}.
For the disk, we have the constraint $a\geq b-k$, and the operators $\hat p^{a,b}_{b-a}$ {\it still} realize an SU($n$)-subalgebra.
Therefore, the invariance of the ZM space under the infinite-dimensional algebra of zero-mode generators implies, {\it in particular}, its invariance under an SU($n$)-subalgebra.
In view of the intimate connection between the ZM generators $\hat p^{a,b}_k$ and the edge effective theory, it is not surprising that this SU($n$) structure has long been associated to Jain CF states based on field theoretic grounds and/or  variational constructions \cite{Read-CF-OP-paper}. Through the present work, this structure becomes an exact feature of a solvable microscopic model for the Jain CF phases (though will of course not remain exact under generic perturbations). For the special case $n=2$, the similarity with the findings of Ref. \cite{Sumanta18} strongly suggests that much of the formalism presented here can be carried over to a rich class of ``parton-like'' states \cite{Jain, Wen90, Wen90_1}, which offer a large playground for the exploration of non-Abelian topological phases \cite{Mostafatobepublished}. We leave this as an interesting challenge for future work.

\textit{Conclusions.$-$}
The theory of the FQH effect traditionally rests on two pillars: (i) quantum-many-body wave functions and (ii) effective field theories. Hamiltonians that are exactly solvable {\it and} fall into the ZM paradigm 
provide a transparent connection between these pillars. The incredibly detailed link between the microscopics and effective field theory provided by edge mode counting has no counterpart in any other area of strongly correlated physics in more than one dimension. Even among the myriad phases of the FQH regime, the definitive parent Hamiltonians satisfying this paradigm cannot always be given. This used to be the situation for the most important class of phases in this regime, those described by Jain CF states. The present work exposes the underlying reasons for this and solves this problem by departing considerably from traditional Hamiltonian construction principles. The latter seek to describe a suitable few-body density matrix via analytic clustering principles. This cannot be done adequately in the case at hand. Instead, we circumvent this problem by an algebraic characterization of few-body correlations in a suitable operator framework. Apart from giving a satisfying solution to the lack of parent Hamiltonians for Jain states  \cite{note11}, we expect the formalism presented here to be of profound value in the exploration of vast classes of more complicated mixed-LL wave functions realizing rich non-Abelian physics, as well as to complement traditional lowest-LL methods. We are hopeful that this angle will inspire exciting future developments.  
\begin{acknowledgments}
The work of S.B. was supported by the VILLUM FONDEN via the Centre of Excellence for Dirac Materials (Grant No. 11744) and the European Research Council under the European Union's Seventh Framework ERS-2018-SYG 810451 HERO. G.O. would like to acknowledge partial support by the DOE Grant DE-SC0020343 and the Clark Way Harrison Professorship at Washington University.
We would like to thank J.K. Jain for insightful discussions.
\end{acknowledgments}
\bibliography{RF}

\begin{thebibliography}{54}
\bibitem [{not()}]{note11}%
  \BibitemOpen
  \href@noop {} {}\bibinfo {note} {While we focus on fermions here for brevity,
  generalization of our results to bosons, ${\sf p}$ half-odd integer, is
  straightforward.}\BibitemShut {Stop}%
\bibitem [{jus()}]{justify}%
  \BibitemOpen
  \href@noop {} {}\bibinfo {note} {To justify the term ``excitation'', a
  confining potential that goes as some power of total angular momentum could
  be added.}\BibitemShut {Stop}%
\bibitem [{com()}]{comm2}%
  \BibitemOpen
  \href@noop {} {}\bibinfo {note} {The Hermitian conjugates $\tilde
  c_{a,j}^\dagger$ appearing in this expression are related to the
  pseudoconjugates $\tilde c_{a,j}^\ast$ via $\tilde c_{a,j}^\dagger=(\tilde
  c_{a,j})^\dagger= A^{-1}_{ca}A^{-1}_{cb} \tilde c^\ast_{b,j}$.}\BibitemShut
  {Stop}%
\bibitem [{foo()}]{footnote1}%
  \BibitemOpen
  \href@noop {} {}\bibinfo {note} {We assume $\ket{\psi}$ to be an eigenstate
  of both angular momentum as well as $\mathfrak{R}$, as these operators
  commute with one another and with $H_{n,\sf p}$.}\BibitemShut {Stop}%
\bibitem [{poi()}]{point}%
  \BibitemOpen
  \href@noop {} {}\bibinfo {note} {This is still true for the half-infinite
  cylinder, where the constraint is $k\geq0$.}\BibitemShut {Stop}%
\end{thebibliography}
%\end{document}
\widetext
\clearpage
\begin{center}
\end{center}
\setcounter{equation}{0}
\setcounter{figure}{0}
\setcounter{table}{0}
\setcounter{page}{1}
\makeatletter
\renewcommand{\theequation}{S\arabic{equation}}
\renewcommand{\thefigure}{S\arabic{figure}}
\renewcommand{\bibnumfmt}[1]{[S#1]}
\renewcommand{\citenumfont}[1]{S#1}
\pagenumbering{roman}
\begin{center}
\textbf{\large Supplemental Material: Local two-body parent Hamiltonians for the entire Jain sequence}
\end{center}
\section{Locality of the Hamiltonian}

Here, we establish that Hamiltonians in the general class discussed in the main text can be chosen to be local {\it in spatial coordinates}. (As remarked in the Introduction, it is not strictly local in the orbital ``lattice'' referred to by the $c_{m,\ell}^{\;}$, $c_{m,\ell}^\dagger$ operators.) This gives us opportunity to make contact 
with the first quantized picture of the Hamiltonian, and elaborate further why a first quantized definition is prohibitive (much unlike other familiar Hamiltonians in the field). We begin by slightly formalizing the setting of our main text. Let ${\cal L}_m$ be the single particle Hilbert space of the $(m+1)$th LL. We work with single particle spaces
\be
   {\cal H}_n= \bigoplus_{m=0}^{n-1} {\cal L}_m\;,
\ee
which define the Fock spaces on which the Hamiltonians  $H_{n,{\sf p}}$ of the main text act.
In the following, we will be particularly interested in the 2-particle spaces $\bigwedge^2 {\cal H}_n$. (We specialize again to fermions for brevity.) We will work in disk geometry here. 
If we consider the limit $n\rightarrow\infty$ (i.e., that including all LLs), we may give the familiar decomposition of $\bigwedge^2 {\cal H}_\infty$ into 2-particle subspaces ${\cal H}_{j,j_r}$ of well-defined total angular momentum $j$ and relative angular momentum $j_r$:
\be\label{decom}
\bigwedge\nolimits^{\!2} {\cal H}_\infty=\bigoplus_{j,j_r} {\cal H}_{j,j_r}\,.
\ee
These subspaces ${\cal H}_{j,j_r}$ are spanned by a basis with (un-normalized) wave functions of the form (where the coordinates are measured in units of the magnetic length $l_B=1$, and $\hbar=1$),
\be \label{Jbasis}
    \psi_{j,j_r,\alpha,\beta}=
    (\bar z_1-\bar z_2)^\alpha   
    (z_1-z_2)^{j_r+\alpha}    
    (\bar z_1+\bar z_2)^\beta
    (z_1+ z_2)^{j+\beta-j_r}\,
    e^{-\frac 18|z_1-z_2|^2-\frac 18 |z_1+z_2|^2 } \quad\mbox{($j_r$ odd for fermions)},
\ee
with $\alpha,\beta\geq 0$, $j_r\geq-\alpha$, and $j\geq j_r-\beta$.
On the other hand, it is important to appreciate that for {\it finite} $n>1$, the spaces
$\bigwedge\nolimits^{\!2} {\cal H}_n$ {\it cannot} be given a basis of the form \eqref{Jbasis}: While for small enough $\alpha$, $\beta$, the states ${\psi_{j,j_r,\alpha,\beta}}$ will be {\it contained} in $\bigwedge\nolimits^{\!2} {\cal H}_n$, there are always those ${\psi_{j,j_r,\alpha,\beta}}$ that are neither contained in $\bigwedge\nolimits^{\!2} {\cal H}_n$ nor in its orthogonal complement (defined in $\bigwedge\nolimits^{\!2} {\cal H}_\infty$). Related to that, the $\bigwedge\nolimits^{\!2} {\cal H}_n$ are not invariant subspaces of the relative angular momentum operator. Now, since $\bigwedge\nolimits^{\!2} {\cal H}_n \subset \bigwedge\nolimits^{\!2} {\cal H}_\infty$, any ${\cal T}\in \bigwedge\nolimits^{\!2} {\cal H}_n$ does have a wave function expansion of the form
\be \label{relmom}
    {\cal T}= \sum_{j,j_r,\alpha,\beta} c_{j,j_r,\alpha,\beta}\, \psi_{j,j_r,\alpha,\beta} \,,
\ee
where, however, some terms on the righthand side may not be in $\bigwedge\nolimits^{\!2} {\cal H}_n$ (though, of course, their components in the orthogonal complement will cancel). These observations are intimately tied to the underlying reason why, unlike in the case $n=1$ \cite{haldane_hier}, the Hamiltonians defined in the main text cannot be easily characterized in terms of relative angular momentum, or, more generally, clustering properties. We will still make use of the general expansion \eqref{relmom} in the following.

We now first turn to the operators $T^r_{a,b,J}$ of the main text and make their spatial dependence more explicit. We have
\be
   T^{r \dagger}_{a,b,J} =
   \int dz_1dz_2\, {\cal T}(z_1,\bar z_1, z_2,\bar z_2)\, \hat\psi^\dagger(z_1)\hat\psi^\dagger(z_2)\,,
\ee
where $\hat\psi^\dagger(z)$ creates an electron localized at
$z$, and ${\cal T}(z_1,\bar z_1, z_2,\bar z_2)$ is the wave function of the 2-particle state created by $T^{r \dagger}_{a,b,J}$ (whose dependence on $r,a,b,J$ we leave understood for the moment). The expression of
$T^{r \dagger}_{a,b,J}$ given in the main text (Eq. (7)) consists of a finite number of terms, so ${\cal T}(z_1,\bar z_1, z_2,\bar z_2)$ is a polynomial (up to a Gaussian prefactor multiplying it), and its expansion of the form  \eqref{relmom} thus has a finite number of terms.
Indeed, we can make that expansion explicit by Taylor expanding $\cal T$ in the new variables $z_1\pm z_2$, $\bar z_1\pm \bar z_2$. For a given number ($n$) of LLs, powers in $\bar z_1$, $\bar z_2$ are at most $\leq n-1$, and the aforementioned 
Taylor expansion then implies
that $\alpha, \beta\leq 2(n-1)$ in \eqref{relmom}, while $j=2J$ is fixed. In particular, $\alpha\leq 2n-2$ gives $j_r\geq -2n+2$. Since we are interested in the dependence on relative coordinates, we will now show that there is also an upper bound on the $(z_1-z_2)$-exponent   (($j_r+\alpha$) of \eqref{Jbasis}) that, in particular, is independent of $j=2J$.

The $T^{r }_{a,b,J}$ annihilate, in particular, all 2-particle zero modes $\ket{\zeta}$ in Fock space, such that $\braket{{\sf vac} | T^{r }_{a,b,J} |\zeta}=0$, equivalently, $\braket{\zeta | T^{r \dagger}_{a,b,J} | {\sf vac}}=0$, i.e., $T^{r \dagger}_{a,b,J} | {\sf vac}\rangle$ is orthogonal to all 2-particle zero modes, and $\ket{\sf vac}$ is the 0-particle vacuum state.
We will utilize this to gain further insight into the center-of-mass/relative angular momentum decomposition \eqref{relmom} of the states
$T^{r \dagger}_{a,b,J} | {\sf vac}\rangle$.
To this end, consider the following alternative basis of ${\cal H}_{j,j_r}$:
\begin{equation}\label{Lbasis}
\begin{split}
   \phi_{j,j_r,\alpha,\beta} &= \hat B_-^{\dagger \alpha}\, \hat A_- ^{\dagger j_r+\alpha} \,\hat B_+^{\dagger \beta}\,\hat A_+^{\dagger j+\beta-j_r}  \,e^{-\frac 18|z_1-z_2|^2-\frac 18 |z_1+z_2|^2 }\quad\mbox{($j_r$ odd for fermions)},\\
   \hat A^\dagger_{\pm}&=\frac 14 (z_1\pm z_2) -(\partial_{\bar z_1}\pm\partial_{\bar z_2}) ,\\
   \hat B^\dagger_{\pm}&=\frac 14 (\bar z_1\pm \bar z_2) -(\partial_{z_1}\pm\partial_{ z_2})\,,
\end{split} 
\end{equation}
The operators $\hat A_\pm ^\dagger $, $\hat B_\pm ^\dagger $, and their Hermitian adjoints are the usual LL ladder operators
with ``$+$'' referring to center-of-mass, ``$-$'' to relative coordinates, $\hat A$ referring to guiding center, and $\hat B$ to dynamical momenta.
That the quantum numbers $j$ and $j_r$ have the same meaning as in \Eq{Jbasis} can be seen by expressing the 
angular momentum operator as $\hat L=\hat L_{\sf cm}+\hat L_{r}$, with $\hat L_{\sf cm}=\hat A_+^\dagger \hat A_+^{\;} - \hat B_+^\dagger \hat B_+^{\;}$,
$\hat L_{r}=\hat A_-^\dagger \hat A_-^{\;} - \hat B_-^\dagger \hat B_-^{\;}$, the latter being center-of-mass and relative contributions, respectively.
Expressing complex coordinates in \Eq{Jbasis} through ladder operators, one easily verifies the relation
\begin{equation}\label{dominant}
   \psi_{j,j_r,\alpha,\beta}\, \sim \,\phi_{j,j_r,\alpha,\beta}  + \mbox{``subdominant'',}
\end{equation}
where $\sim$ implies a non-zero proportionality factor, and ``subdominant'' refers to a linear combination of terms in the basis  \Eq{Lbasis} with powers in the operators 
$\hat A^\dagger_\pm$, $\hat B^\dagger_\pm$ that are all less than or equal to corresponding powers in the leading term (with at least one being strictly less).  
%Considering now the analog of \Eq{relmom}
 %\be \label{relmom2}
 %   {\cal T}= \sum_{j,j_r,\alpha,\beta} \tilde c_{j,j_r,\alpha,\beta}\, \phi_{j,j_r,\alpha,\beta} \,,
%\ee
%by means of \Eq{dominant}, the bounds $\alpha,\beta\leq 2(n-1)$ we had for \Eq{relmom} immediately also applies to \Eq{relmom2}. 
 
From \Eq{Jbasis}, it is clear that any element of ${\cal H}_{j,j_r}$ with $j_r\geq 2\sf p$ contains a factor of $(z_1-z_2)^{2\sf p}$. Therefore, 
any element of  $\bigwedge^2 {\cal H}_n$ whose expansion \Eq{Jbasis} only contains terms with $j_r\geq 2\sf p$ will be, by the results of the main text, 
a zero mode of $H_{n,\sf p}$. (Note, however, that not all zero modes need to have $j_r\geq 2\sf p$ !) We can obtain such a zero mode by starting with any $\ket{\xi} \in \bigwedge^2 {\cal H}_n$, and applying
$\hat A_-^{\dagger \,(2{\sf p} +2n-2)}$. Towards this end we may recall that (i) $j_r\geq -2n+2$, note that (ii) $\hat A_-^{\dagger \,(2{\sf p} +2n-2)}$ increases $j_r$ 
by $(2{\sf p} +2n-2)$, and furthermore note that (iii) $(\hat A_-^\dagger)^2$ is a well-defined operator acting on $\bigwedge^2 {\cal H}_n$  (it only {\it lowers} powers of $\bar z_1$, $\bar z_2$ and preserves the parity of $j_r$), $\hat A_-^{\dagger \,(2{\sf p} +2n-2)}\ket{\xi}$ is an element of $\bigwedge^2 {\cal H}_n$ with zero amplitude for $j_r<2\sf p$. Putting all of these pieces together one indeed verifies that $\hat A_-^{\dagger \,(2{\sf p} +2n-2)} \ket{\xi}$ is a zero mode. Writing $\ket{\cal T} = T^{r \dagger}_{a,b,J}\ket{\sf vac}$, as explained above, $\ket{\cal T}$ is orthogonal to all zero modes, thus,
\begin{equation}\label{ortho}
         \braket{{\cal T} | \hat A_-^{\dagger \,(2{\sf p} +2n-2)}|\xi}= 0=\braket{{\xi} | \hat A_-^{ \,(2{\sf p} +2n-2)}|\cal T}\;\mbox{for all }\ket{\xi}\in \sideset{}{^2}\bigwedge {\cal H}_n\;.
\end{equation}
One similarly notes that $(\hat A_-)^2$ leaves 
$\bigwedge^2 {\cal H}_n$ invariant: One only needs to observe that $[\hat A_-,  \bar z_i(\partial_{\bar z_i} +\frac 14 z_i  )]=0$, i.e.,
$\hat A_-$ commutes with the operator that counts the degree of the polynomial part of the wave function in $\bar z_i$, $i=1,2$.
Then \Eq{ortho}, for all $\ket{\xi}\in \bigwedge^2 {\cal H}_n$, implies that $\hat A_-^{ \,(2{\sf p} +2n-2)}\ket{\cal T}=0$.
In the expansion \Eq{relmom} of the wave function $\cal T$ of $\ket{\cal T}$, consider now the leading non-zero term by lexicographical order 
  in $(j_r+\alpha,\alpha , \beta)$ ($j=2J$ is fixed). In particular, this term has maximum $j_r+\alpha$. 
  Let $\tilde\alpha, \tilde\beta,\tilde j_{r}$ be the quantum numbers of this term. By \Eq{dominant}, this leading term is the only
  term making a non-zero contribution to $\phi_{j,\tilde j_{r},\tilde \alpha,\tilde \beta}$  when switching to the basis \eqref{Lbasis}.
  Assume, now, that $\tilde j_{r}+\tilde\alpha \geq2{\sf p} +2n-2$. Then $\hat A_-^{ \,(2{\sf p} +2n-2)}\ket{\cal T}$ would 
  have non-zero overlap with $\hat A_-^{ \,(2{\sf p} +2n-2)}   \phi_{j,\tilde j_{r},\tilde \alpha,\tilde \beta}   \sim \phi_{j- 2{\sf p} -2n+2, \tilde j_r- 2{\sf p} -2n+2, \tilde \alpha,\tilde\beta}$. Hence $\hat A_-^{ \,(2{\sf p} +2n-2)}\ket{\cal T}$ 
  would not vanish, a contradiction. This proves that for $\ket{\cal T} = T^{r \dagger}_{a,b,J}\ket{\sf vac}$, all terms in the expansion
  \eqref{relmom} have $(z_1-z_2)$-exponents that are less than  $2{\sf p} +2n-2$.

Next we consider the full Hamiltonian of Eq. (6) in the main text. In what follows, we fix the values of the indices $r,a,b$ and omit these to avoid a cumbersome notation. We will focus on a single term of the form 
\be\label{HMnS}
   \sum\limits_J
  E_J \, T^{\dagger}_JT_{J}^{\;}= \int dz_1dz_2dz_3dz_4\, K(z_1,z_2,z_3,z_4) \, \hat\psi(z_1)^\dagger \hat\psi(z_2)^\dagger\hat \psi(z_4)\hat \psi(z_3). 
\ee
We wish to demonstrate that positive coefficients
$E_J$ can always be chosen such that the resulting operator is local.

According to the above, the kernel has the following Gaussian/polynomial structure:
\be
\begin{split}
&K(z_1,z_2,z_3,z_4) =\sum_{j\geq j_{\sf min}} E_{j/2} \sum_{\alpha,\beta,\alpha',\beta'=0}^{2(n-1)}\sum_{j_r=-\alpha}^{\min(2{\sf p}+2n-3-\alpha,j+\beta)}\,\sum_{j_r'=-\alpha'}^{\min(2{\sf p}-1+2n-3-\alpha',j+\beta')} \!\!\!\!\!\!\!\!\!\!
\!\!\!\!\!\!\!\!\!\gamma_{\alpha,\beta,j_r}(j)
\bar \gamma_{\alpha',\beta',j_r'}(j)
(\bar z_1-\bar z_2)^\alpha   
    (z_1-z_2)^{j_r+\alpha}    
    (\bar z_1+\bar z_2)^\beta\\ 
   & (z_1+ z_2)^{j+\beta-j_r}
    ( z_3- z_4)^{\alpha'}
    (\bar z_3-\bar z_4)^{j_r'+\alpha'}    
    ( z_3+ z_4)^{\beta'} 
    (\bar z_3+ \bar z_4)^{j+\beta'-j_r'}
 \,   \exp[-\frac 18(|z_1-z_2|^2+ |z_1+z_2|^2+ |z_3-z_4|^2+ |z_3+z_4|^2)],
\end{split}
\ee
with $\gamma_{\alpha,\beta,j_r}(j):=c_{\alpha,\beta,j,j_r}$
being the polynomial coefficients describing $T_J^\dagger$ according to \Eq{relmom} and $j_{\sf min}=-2n+2$.
This then yields % the estimate
\be\label{est1}
\begin{split}
&|K(z_1,z_2,z_3,z_4)|\leq\sum_{j\geq j_{\sf min} }E_{j/2} \sum_{\alpha,\beta,\alpha',\beta'=0}^{2(n-1)}\sum_{j_r=-\alpha}^{\min(2{\sf p}+2n-3-\alpha,j+\beta)}\,\sum_{j_r'=-\alpha'}^{\min(2{\sf p}-1+2n-3-\alpha',j+\beta')}
\;|\gamma_{\alpha,\beta,j_r}(j)|
| \gamma_{\alpha',\beta',j_r'}(j)|\\
&|z_1- z_2|^{j_r+2\alpha}  |z_1+ z_2| ^{j+2\beta-j_r}
    |z_3- z_4|^{j_r'+2\alpha'}
    |z_3+ z_4|^{j+2\beta'-j_r'}
 \,   \exp[-\frac 18(|z_1-z_2|^2+ |z_1+z_2|^2+ |z_3-z_4|^2+ |z_3+z_4|^2)].
\end{split}
\ee
Next, we choose
\be
  0<E_{j/2}\leq \frac{1}{4^{(j+j_{\sf min})}(j+j_{\sf min})!}\, 
  \frac{1}{(\max_{\alpha,\beta,j_r}|\gamma_{\alpha,\beta,j_r}(j)| )^2}\,.
\ee
If we now take the $j$-sum first, at fixed $\alpha,\beta, j_r$ and $\alpha',\beta', j_r'$ we find
%such that for the $j$-sum (over half-integer values) we have %the estimate
\be\begin{split} \sum_{j\geq j_{\sf min}'}  E_{j/2} |\gamma_{\alpha,\beta,j_r}(j)|
| \gamma_{\alpha',\beta',j_r'}(j)|
|z_1+z_2|^{j+2\beta-j_r}|z_3+z_4|^{j+2\beta'-j_r'}\\
\leq |z_1+z_2|^{j_{\sf min}'+2\beta-j_r}|z_3+z_4|^{j_{\sf min}'+2\beta'-j_r'} \sum_{k=0}^\infty 
\frac{1}{4^k k!} (|z_1+z_2| |z_3+z_4| )^k\\
= |z_1+z_2|^{j_{\sf min}'+2\beta-j_r}|z_3+z_4|^{j_{\sf min}'+2\beta'-j_r'}  \exp[\frac 14  |z_1+z_2||z_3+z_4|]\,.
\end{split}\ee 
where  $j_{\sf min}'\geq j_{\sf min}$ is the lowest $j$-value for given  $\alpha,\beta, j_r$ and $\alpha',\beta', j_r'$ that renders the $\gamma$-factors non-zero, and in particular ensures that the powers in the last line are non-negarive.
 Inserting the above in \Eq{est1} gives
\be\label{est2}
 |K(z_1,z_2,z_3,z_4)|\leq
 {\sf polynomial}\left(|z_1-z_2|,|z_1+z_2|, |z_3-z_4|,|z_3+z_4|\right)\, 
 \exp[-\frac 18 (|z_1+z_2|-|z_3+z_4|)^2-\frac 18 |z_1-z_2|^2 - \frac 18 |z_3-z_4|^2 ]\,,
\ee 
where the degree of the polynomial in the various arguments is bounded by simple expressions in $n$ and $\sf p$ as discussed above.
\Eq{HMnS} then manifestly is a local interaction. 

\begin{figure}
    \centering
    \includegraphics[width=12cm]{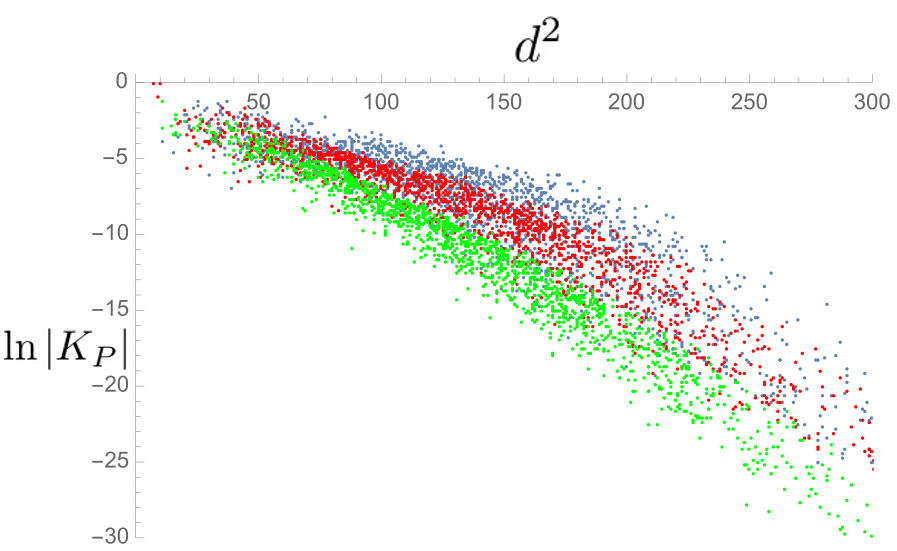}
    \caption{(Color online) Spatial dependence of the kernel $K_P$ for the unique projection operator defined in the text. Vertical axis is the logarithm of the modulus of the kernel. The horizontal axis is the squared distance measure $d^2:=|z_1-z_4|^2+|z_2-z_3|^2+|z_1-z_3|^2+|z_2-z_4|^2$. Shown are the cases $n=2$, ${\sf p}=1$ (green), $n=3$, ${\sf p}=1$  (red), and $n=3$, ${\sf p}=2$ (blue). For each case, $\sim 1500$ random points $(z_1,z_2,z_3,z_4)$ with all $z_i$ within the disk of radius $6$ are plotted. The (super)-exponential decay of the kernel with distance is evident in all cases.}
    \label{KPfig}
\end{figure}

Even at fixed $n$ and $\sf p$, our results apply to a large class of Hamiltonians. This is so since we may not only choose the constants $E_j$ within certain bounds (to preserve locality) but, as we point out in the main text, we may also replace the operators 
$T^r_{a,b,J}$ with any suitable new (independent) linear combinations. In practice, one will often want to work with translationally invariant as well as local Hamiltonians. There is a canonical choice for a translationally invariant Hamiltonian: given any pair $n,{\sf p}$, the class of Hamiltonians defined here contains exactly one member that is a 2-particle projection operator. It is obtained by ortho-normalizing, at each $J$, the 2-particle states created by the $T^{r\dagger}_{a,b,J}$, and forming the corresponding new linear combinations of these operators. We denote the resulting Hamiltonian by $P_{n,{\sf p}}$. When acting on 2-particle states, $P_{n,{\sf p}}$ is the orthogonal projection onto the subspace of general 2-particle CF states. This subspace is translationally invariant, implying that $P_{n,{\sf p}}$ is also translationally invariant. Specifically,
\begin{equation}\label{canonicalprojector}
    P_{n,{\sf p}}=  \sum\limits_{ J, r, a, b,r',a',b'}
  G_{r,a,b,r',a',b'}(J) T^{r\dagger }_{a,b,J}T^{r'}_{a',b',J}\,,
\end{equation}
where for each $J$, $G(J)$ can be viewed as a matrix with multi-indices $\lambda=(r,a,b)$, $\lambda'=(r',a',b')$, which is the inverse of the matrix
\begin{equation}
\begin{split}
    & G^{-1}_{\lambda,\lambda'}(J)=\langle 0|T_{\lambda,J}T_{\lambda',J}^\dagger|0\rangle =\\ &\sum_{x,c,d}x^{r+r'}\left( A^{-1}_{c,b}(J-x) A^{-1}_{c,b'}(J-x)A^{-1}_{d,a}(J+x) A^{-1}_{d,a'}(J+x)-(-1)^{r'} A^{-1}_{c,a}(J+x) A^{-1}_{c,b'}(J+x)A^{-1}_{d,b}(J-x) A^{-1}_{d,a'}(J-x)\right)\,.
\end{split}
\end{equation}
The inverse of the above exists thanks to the linear independence of the states,
$T^{r\dagger }_{a,b,J}\ket{0}$, which (in disk geometry) only requires caution for small $J$: 
We must refine the requirement $0\leq r<2{\sf p}$ from the main text to read
$0\leq r\leq \min(2{\sf p}-1, 2J+a+b)$ in all multi-indices $\lambda$. As before, we also enforce  $0\leq a\leq b<n$, and that $r$ is odd for $a=b$.
\Eq{canonicalprojector} can then also be expressed in the form given on the righthand side of \Eq{HMnS}
with a kernel $K_P(z_1,z_2,z_3,z_4)$. 
We have numerically investigated the locality of this kernel for some choices of $n, {\sf p}$, finding it to be exponentially decaying in a squared-distance-measure as suggested by \Eq{est2}. The results 
%of calculations demonstrating the locality of $K_P$ in disk geometry 
are shown in Fig. \ref{KPfig}.
Table \ref{table1} provides explicit forms for the required $A$-matrices in various geometries. 

\begin{table}[]
    \begin{tabular}{c|c|c } 
        \hline\hline  Geometry & Disk & Cylinder\\\hline
         $A(\ell)_{a,m}$ & $\frac{2^{a}\sqrt{2^{\ell+1} \pi}a!(a+\ell)!}{(a-m)!\sqrt{m!(m+\ell)!}}$& 
            $\sqrt{\frac{2\pi}{\kappa}}{e^{\frac{1}{2}\kappa^2\ell^2}}\frac{\sqrt{\sqrt{\pi}2^mm!}}{2^a}\binom{a}{m}\tilde H_{a-m}(\kappa \ell)$
        \\\hline
         $A^{-1}(\ell)_{m,a}$ & $\frac{(-1)^{m+a}}{\sqrt{2^{\ell+1}\pi}2^{a}}\frac{\sqrt{m!(m+\ell)!}}{a!(m-a)!(a+\ell)!}$ & 
        $\sqrt{\frac{\kappa}{2\pi}}\frac{e^{-\frac{1}{2}\kappa^2\ell^2}}{\sqrt{\sqrt{\pi}2^mm!}}\binom{m}{a}2^{a}H_{m-a}(-\kappa \ell)$\\ [1ex] 
 \hline
    \end{tabular}
    \caption{ Matrix elements for the lower triangular matrices $A(\ell)$,  $A(\ell)^{-1}$  are shown for the disk and cylinder geometries. In both geometries, the matrix $A^{-1}(\ell)_{m,a}$ Gram-Schmidt-orthogonalizes monomial orbitals with given (angular) momentum quantum number $\ell$ to form proper Landau-level states. These monomials are $\bar z^a z^{a+\ell}$ for the disk geometry, and $x^{a}\xi^\ell$ for the cylinder, where $\xi=\exp(\kappa z)$, and $\kappa=2\pi l_B/L_y$ is the inverse cylinder radius in units of the magnetic length    $l_B$. As the cylinder has no shift between $\ell$ and the second monomial exponent (related to guiding center), we would not distinguish between $\tilde c$- and $\tilde d$-operators for the cylinder, unlike we did for the disk in the main text. Rather, $\tilde d:=\tilde c$ for the cylinder. The resulting definition of root states is then the same in the monomial basis for both geometries. $H_m(x)$ is a Hermite polynomial, and $\Tilde{H}_m(x)=\sum\limits_{\mathfrak{a}=0}^{\left \lfloor{m/2}\right \rfloor }2^\mathfrak{a}\frac{(2\mathfrak{a})!}{\mathfrak{a}!}\binom{m}{2\mathfrak{a}}H_{m-2\mathfrak{a}}(x)$.}
    \label{table1}
\end{table}

\begin{table}[]
{\renewcommand{\arraystretch}{1.1}
    \begin{tabular}{|@{\hspace{1em}}c@{\hspace{1em}}|@{\hspace{1em}}c@{\hspace{1em}}|@{\hspace{1em}}c@{\hspace{1em}}|@{\hspace{1em}}c@{\hspace{1em}}|c |} 
        \hline  $n$ & $2{\sf p}$ & $N$ &$j$& \# of zero modes\\\hline
       3      &         2    &           3      &         1         &      0   \\
3      &         2     &          3      &         2          &     1\\
3      &         2      &         3       &        3        &       4\\
3      &         2      &         3       &        4        &       12\\
3      &         2      &         3       &        5        &       24\\
3      &         2       &        4        &       7         &      0\\
3       &        2        &       4         &      8         &      3\\
3        &       2     &          4     &          9        &       9\\
3         &      2      &         4      &         10       &      25\\
3          &     2       &        4       &        11       &      47 \\
3     &          2       &        5        &       15      &        0\\
3     &          2        &       5         &      16        &      3\\
3      &         2     &          5      &         17        &      11\\
3       &        2      &         5      &         18        &      31\\
3      &         2       &        5       &        19         &     66\\
2       &        4        &       5      &         40        &      0\\
2      &         4      &         5        &       41        &      1\\
2      &         4       &        5        &       42        &      4\\
 \hline
    \end{tabular}}
    \caption{Number of zero modes from exact diagonalization of \Eq{canonicalprojector} at given $n$, $\sf p$, particle number $N$ and angular momentum $j$ in disk geometry. The last column agrees with \Eq{ncf}, the number of CF states with given $n$, $\sf p$, $N$, and $j$, in all cases examined.}
    \label{table2}
\end{table}
To test the main results of this Letter, we have diagonalized \Eq{canonicalprojector} for various values of $n$, $\sf p$, particle number $N$, and in various sectors of angular momentum $j$. Table \ref{table2} shows the number of zero modes found in each case. These numbers may be compared with the number of CF states in disk geometry having the same quantum numbers, calculated from the formula
\begin{equation}\label{ncf}
n_{\sf CF} (n,{\sf p}, N, j )= \sum_{\substack{N_1,\dotsc_,N_n\\N_1+\dotsc+N_n=N\\N_i\geq 0}}\;\;
\sum_{\substack{j_1,\dotsc_,j_n\\j_1+\dotsc+j_n=j-{\sf p}N(N-1)\\j_i\geq -\frac 12 i(i-1)}}
\prod_{i=1}^n  \;q(j_i+iN_i, N_i)\,,
\end{equation}
where $q(j,N)$ denotes the number of partitions of the integer $j$ into exactly $N$ positive integer parts without repetition (and in particular gives $0$ if  $j<0$, or if $j>0, N=0$, or if $j=0, N>0$, and furthermore $q(0,0):=1$).
We found agreement with \Eq{ncf} for all cases listed in Table \ref{table2}.

\section{Entangled Pauli principle: detailed derivation and completeness of composite fermion zero modes}

Here, we will give rigorous derivations of the entangled Pauli principle (EPP) and its main consequence, the completeness of the CF states as zero modes of their respective parent Hamiltonian. In keeping with our definition of root states given in the main text, we will work with the operators $\tilde d_{a,\fr}$, $\tilde d_{a,\fr}^\ast$ given there. 
Note that the root states associated to the EPP necessarily agree with the state's thin cylinder limit, which is further identical (modulo boundary conditions) to the thin torus limit \cite{BK,BK1,BK2,BK3,seidel05,seidel2006abelian,SY1,Seidel10,SY2,FlavinPRX,Flavin12,zhou2013heat} (cf. also the caption of Table \ref{table1} in this context). It is furthermore beneficial to introduce the operators  
%It is easy to see that the $Q^r_{a,b,\fr}$ are (linearly independent) linear combinations of operators $T^r_{a,b,J}$ with $J=\fr-(a+b)/2$ defined in the main text. 
\begin{equation}
    Q^r_{a,b,\fr}=\sum_{x} x^r \tilde{d}_{a,\fr+x}\tilde{d}_{b,\fr-x}\,.
\end{equation}
It is easy to see that $Q^r_{a,b,\fr}$ are linearly independent combinations of the operators $T^{r'}_{a',b',J}$ defined in the main text with $J=\fr-(a+b)/2$. The zero mode condition for a ket $\ket{\psi}$ therefore implies (and is equivalent to)
\begin{equation}
    Q^r_{a,b,\fr}\ket{\psi}=0 \quad\mbox{for all $r,a,b,\fr$ ($\fr$ integer or half-odd-integer).}
\end{equation}

We will first
derive a quintessential aspect of the EPP - that forbidding more than a single occupancy at any given $\fr$-value at the root level. We will establish this via a proof by contradiction. Towards this end, we let $\ket{\psi}$ be a zero mode and, omitting normalization factors, write it as
\be\label{root}
\ket{\psi} = \tilde d^\ast_{b,\fr} \tilde d^\ast_{a,\fr}\ket{S}+ \sum_{S'}\ket{S'} 
\ee 
where $a\neq b$, $\ket{S}$ is a $N-2$-particle Slater determinant with definite occupancies in the $\tilde d^\ast$-basis, and with the  $d^\ast_{a,\fr}$, $d^\ast_{b,\fr}$ states unoccupied. The sum over $S'$ is over $N$-particle Slater-determinants from the same basis, all having occupancies differing from the first term. All individual Slater determinants contain implicit phase and normalization factors. Let us now assume that the first term is non-expandable \cite{ortiz2013} (as defined in the main text) thus contributing to the root state. 
Since $\ket{\psi}$ is a zero mode, 
\be \label{ZMC}
0=Q^0_{a,b,\fr}\ket{\psi} =\ket{S} +\sum_{x\neq 0} \tilde d^\ast_{b,\fr} \tilde d^\ast_{a,\fr} 
\tilde{d}_{a,\fr+x}\tilde{d}_{b,\fr-x}\ket{S}
+\sum_{S'} \sum_x \tilde{d}_{a,\fr+x}\tilde{d}_{b,\fr-x} \ket{S'}\,.
\ee
Each term in the above is, evidently, a Slater determinant, and all Slater determinants must cancel. In the first sum, every term is manifestly a Slater determinant different from $\ket{S}$. By `different', we mean a different member, up to a phase, of the set of linearly independent Slater determinants generated by the $\tilde d^\ast$-operators. It thus follows that any term canceling the first term, $\ket{S}$, must originate from the second sum. Suppose then that
\be
  \tilde{d}_{a,\fr+x}\tilde{d}_{b,\fr-x}\ket{S'} \propto \ket{S}\,,
\ee
implying that
\be\label{squeeze}
 \tilde d^\ast_{b,\fr} \tilde d^\ast_{a,\fr} \tilde{d}_{a,\fr+x}\tilde{d}_{b,\fr-x}\ket{S'} \propto \tilde d^\ast_{b,\fr} \tilde d^\ast_{a,\fr} \ket{S}\,.
\ee
For $x=0$, the LHS must vanish (otherwise, we will contradict the assumption that all $\ket{S'}$ have different occupancy configurations than the first term in \Eq{root}). For $x\neq 0$,
\Eq{squeeze} indicates that we can obtain 
the first term in \Eq{root} by an ``inward squeezing'' process (moving two particles closer to each other while preserving angular momentum) applied to the Slater determinant $\ket{S'}$
appearing in the Slater-decomposition of $\ket{\psi}$, or 
conversely,
that we can obtain $\ket{S'}$ from the first term in \Eq{root} by an ``expansion'' process as defined in the main text. This contradicts the assumption that the first term in \Eq{root} was non-expandable.
Thus, no term in the root state can have double occupancies.
 
We may proceed similarly to show that, assuming the first term is non-expandable, no decomposition of the form
\be\label{root2}
\ket{\psi} = \tilde d^\ast_{b,\fr-\Delta/2} \tilde d^\ast_{a,\fr+\Delta/2}\ket{S}+ \sum_{S'}\ket{S'} 
\ee 
is possible, where $0<\Delta<2{\sf p}$ is an even (odd) integer if $\fr$ is an integer (half-odd integer). 
To this end, we consider the operators
\be
{\cal Q}^p_{a,b,\fr}=\sum_{x} p(x) \tilde{d}_{b,\fr+x}\tilde{d}_{a,\fr-x}\,
\ee
where $p(x)$ is a polynomial.
Obviously, so long as the degree of $p$ is less than $2{\sf p}$, the operator ${\cal Q}^p_{a,b,\fr}$ is a linear combination of the operators ${Q}^r_{a,b,\fr}$  (for $a=b$, the even part of $p$ is irrelevant).
For a given $\Delta$, we may construct an even (odd) polynomial $p_e$ ($p_o$) of degree at most $\Delta$ such that $p_{e/o}(\Delta/2)=1$ and $p_{e/o}(x)=0$ for $x=-\Delta/2+1,\dotsc,  \Delta/2-1$. For even $\Delta$, we have
$p_o(x)\propto\prod_{k=-\Delta/2+1}^{\Delta/2-1}(x-k)$ of degree $\Delta-1$, $p_e\propto x p_o$, with roles reversed for $\Delta$ odd. 
We will now first consider the case in which $a\neq b$. By forming $p_+=(p_e+p_o)/2$, we have $p_+(\Delta/2)=1$, $p_{+}(x)=0$ for $x=-\Delta/2,\dotsc,  \Delta/2-1$. There is then no obstruction for applying the reasoning of \Eq{ZMC} to $0={\cal Q}^{p_+}_{a,b,\fr}\ket{\psi}=\dotsc$. Proceeding as before then shows that the first term in \Eq{root2} could not be part of the root state.
When $a=b$, we can simply work with $p_o$. It is noteworthy that this reasoning still applies when $\Delta=2{\sf p}$, since then $p_o$ is of degree $2{\sf p}-1<2{\sf p}$. The last case to be considered is thus that of $a\neq b$ and $\Delta=2{\sf p}$. To this end, we expand
\be\label{root3}
\ket{\psi} =A\, \tilde d^\ast_{a,\fr-{\sf p}} \tilde d^\ast_{b,\fr+{\sf p}}\ket{S}+ B\, \tilde d^\ast_{b,\fr-{\sf p}} \tilde d^\ast_{a,\fr+{\sf p}}\ket{S}+\sum_{S'}\ket{S'} \,,
\ee
where, again, all of the states
$\ket{S'}$ have occupancies differing from those in the first two terms.
Since ${\sf p}$ is integer, $p_o$ has degree $2{\sf p}-1$ and $p_e$ has degree ${2\sf p}$. Thus, only $p_o$ is available. Evaluating $0={\cal Q}^{p_o}_{a,b,\fr}\ket{\psi}=\dotsc$, and using the same arguments as in the above demonstrates that only the first two terms in \Eq{root3} can contribute to $\ket{S}$, yielding $A=-B$ (assuming again that these first two terms are non-expandable, i.e., are root-level terms). Putting all of the pieces together thus proves the EPP of the main text. Valid root states are, therefore, linear combinations of products of clusters of the form
\be\label{rootket}
\includegraphics[scale=.4]{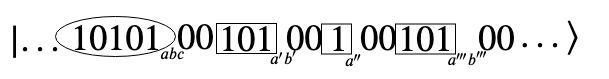}
\ee
where each cluster consists of at most $n$ particles separated by $2{\sf p}$ orbitals (in \Eq{rootket}, ${\sf p}=1$, $n=3$ with ovals denoting an SU($n$) singlet)
and is totally antisymmetric in the $\Lambda$L-indices. In \Eq{rootket}, the $\Lambda$Ls that are occupied in each cluster are indicated by subscripts $0\leq a<b\dotsc <n$, etc..
(could be omitted for
the singlet case [ovals] with $n$ particles).
Different clusters are separated by at least $2{\sf p}+1$ orbitals. It is worth noting that each cluster carries a well-defined value for both $\mathfrak{R}=\sum_{a,\fr} \fr \, \tilde d_{a,\fr}^\ast \tilde d_{a,\fr}$ as well as angular momentum $\hat L$ (where both operators clearly are diagonalizable, but only $\hat L$ is Hermitian). This is of a piece with the observation that $H_{n,{\sf p}}$, Eq. (6) of the main text, commutes with both operators, and we should thus find a basis of zero modes that are both eigenstates of $\hat L$ and $\mathfrak{R}$. We now decompose the Hilbert space as follows,
\be\label{decom1}
{\cal H}= {\cal H}^\sfr \,\oplus \,{\cal H}'\,,
\ee
where ${\cal H}^\sfr$ is spanned by all potential root states of the form \eqref{rootket}, and ${\cal H}'$ is spanned by the following three types of states: i) single Slater determinants with two particles separated by less than $2{\sf p}$, ii) single Slater determinants with more than $n$ consecutive spins separated by $2{\sf p}$ orbitals, iii) product states similar to \eqref{rootket} but where at least one of the clusters is of a different symmetry type (described by a different Young tableau), rather than totally anti-symmetric in $\Lambda$L-indices. Obviously, then, the sum in \Eq{decom1} is direct.  We may also withdraw to subspaces of given total angular momentum, so as to keep the Hilbert spaces appearing in \Eq{decom1} finite dimensional, writing
\be\label{decomL}
{\cal H}_j= {\cal H}^\sfr_j \,\oplus \,{\cal H}'_j\,.
\ee

We can then prove the following, quite general 

\underline {Theorem:} {\it The number of linearly independent zero modes of the Hamiltonian  (Eq. (6) in the main text) at given angular momentum $j$ is at most equal to the dimension of the subspace of ${\cal H}^\sfr_j$.}

Proof: 
Let ${\cal H}^Z_j\subset {\cal H}_j$ be the subspace of zero modes of angular momentum $j$.
Any $\ket{\psi}\in {\cal}H_j^Z$ can be uniquely decomposed as
$\ket{\psi}= \ket{\psi^r}+\ket{\psi'}$
with $\ket{\psi_r}\in {\cal H}_{j}^\sfr$ and $\ket{\psi'}\in {\cal H}_j'$. Assume now that $n_{j}^r:=\mbox{dim} {\cal H}_{j}^\sfr$ is less than $n_{j}^Z:=\mbox{dim} {\cal H}_{j}^Z$.
The linear projection $P: {\cal H}_{j}^Z\rightarrow {\cal H}_{j}^\sfr$, $\ket{\psi} \mapsto \ket{\psi^r}$ must then have non-zero kernel, so there is a nonzero $\ket{\psi}\in {\cal H}_{j}^Z$ with $\ket{\psi^r}=0$. Together with $\ket{\psi}$, the root state of this $\ket{\psi}$ must then also lie in ${\cal H}'_j$, as follows straightforwardly from the
construction of ${\cal H}'_j$ and the definition of the root state.
This violates our EPP, a contradiction $\square$

Lastly, we will now show that, in the notation of the proof, $n^Z_j=n^r_j$ for the models under consideration. As explained in the main text, this establishes a ``zero mode paradigm'' for these models. Recall that, in the main text, we demonstrated that CF states of the form
\be\label{CF2}
  \prod_{1\leq i<j\leq N} (z_i-z_j)^{2\sf p} \times \mbox{``$n-\Lambda$L Slater determinant''}
\ee
are zero modes of $H_{n,\sf p}$. 
We may parametrize Slater determinants by an occupation number matrix $\mathfrak{n}_{a,\fr}$ of $1$s and $0$s, where $0\leq a<n$ and  $\fr\geq 0$. Let $N_\fr:=\sum_{a=0}^{n-1} \mathfrak{n}_{a,\fr}$ be the number of particles in the $\fr$-column of the matrix. 
Assuming $N_\fr>0$, the particles in the $\fr$-column can be associated with a cluster of the form appearing in \Eq{rootket} (see also Fig. (1) and caption), where the beginning orbital of the cluster has $\mathfrak{R}$-index $\tilde \fr= 2{\sf p}\sum_{\fr'<\fr} N_{\fr'}+\fr$, and the terminal orbital has $\mathfrak{R}$-index $\tilde \fr+2{\sf p}(N_\fr-1)$, and the $\Lambda$L-indices occupied in the cluster are precisely given by the non-zero $\mathfrak{n}_{a,\fr}$ ($\fr$ fixed!). It is easy to see that the product of the clusters associated to the Slater-determinant in this way gives a state of the from \eqref{rootket}, i.e., a possible root state, indeed one of the same angular momentum as the associated CF-state \eqref{CF2}. One could show that this product of clusters is indeed {\it the} root state of the associate CF-state. However, this is not necessary here, since it is easy to see that the mapping described here between all possible CF-states \eqref{CF2} and the set of all possible root kets of the form \eqref{rootket} is onto, i.e., for each such root ket, we can construct a Slater-determinant/CF state associated to it. Since all these CF states are linearly independent (as the underlying Slater determinants certainly are), and are zero modes, this proves $n^Z_j\geq n^r_j$.
Since we have already proven the opposite bound above, we must have
\be
n^Z_j =n^r_j\,.
\ee
This concludes the proof that there is a one-to-one correspondence between root ``patterns'' of the form \eqref{rootket}, and a set of linearly independent zero modes.
In particular, the CF-states \eqref{CF2} form a complete set of zero modes of the Hamiltonian $H_{n,{\sf p}}$, which is the chiefly desired property of the construction presented in this paper.

\end{document}